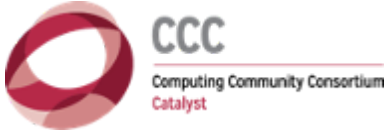

# Pandemic Informatics: Preparation, Robustness, and Resilience
*A Computing Community Consortium (CCC) Quadrennial Paper*

*Elizabeth Bradley (University of Colorado Boulder), Madhav Marathe (University of Virginia), Melanie Moses (The University of New Mexico), William D Gropp (University of Illinois Urbana-Champaign), and Daniel Lopresti (Lehigh University)*

## Overview

Infectious diseases cause more than 13 million deaths a year, worldwide. Globalization, urbanization, climate change, and ecological pressures have significantly increased the risk of a global pandemic. The ongoing COVID-19 pandemic—the first since the H1N1 outbreak more than a decade ago and the worst since the 1918 influenza pandemic—illustrates these matters vividly. More than 47M confirmed infections and 1M deaths have been reported worldwide as of November 4, 2020 and the global markets have lost trillions of dollars. The pandemic will continue to have significant disruptive impacts upon the United States and the world for years; its secondary and tertiary impacts might be felt for more than a decade.

An effective strategy to reduce the national and global burden of pandemics must: (*i*) detect timing and location of occurrence, taking into account the many interdependent driving factors; (*ii*) anticipate public reaction to an outbreak, including panic behaviors that obstruct responders and spread contagion; and (*iii*) develop actionable policies that enable targeted and effective responses. These three aims will require advances in a number of areas, including:

- The development of models that are not just scientifically effective, but that support understanding on the part of the public, as well as actionable insights for policy makers.
- Identification and preparation of computational and data resources (data, computational power, expertise) that will allow us to respond quickly and predict effectively in a crisis situation.
- Real-time collection and updating of data, models, and model assumptions in rapidly changing environments.

These are not purely technological problems. Effective preparation for and response to future pandemics will require integration of solutions that span the full sociotechnical spectrum of challenges that are posed by these devastating events. This will require systemic, national-level support and a coordinated effort by the computing research community, in tandem with a broad coalition of experts from the social and political sciences, economics and the humanities. Such a framework will allow us to develop an understanding across scales, from cells and RNA to epidemic spread through communities and across countries. Only with such a comprehensive understanding will we be prepared to more effectively manage the next pandemic.

**Pandemic Modeling**

The first COVID-19 models communicated the extraordinary rates of illness and death that could result from an uncontrolled pandemic. That understanding led many governments to implement control measures early enough to mitigate the disease's impact and save millions of lives.

However, models are simplifications of a very complex reality. For instance, age, socio-economic status, race, and environmental conditions create disparities in infection and mortality rates, but these realities are often not incorporated in the models that have informed policy in the COVID pandemic. More broadly, few epidemiological models factor in antimicrobial resistance, zoonosis, climate change, or increased urbanization, all of which will play increasingly important roles in future pandemic outbreaks.

Models and modeling frameworks that incorporated these realities could help us more effectively prepare for, and respond to, pandemics—e.g., identifying targeted interventions that will save the most lives, designing methods for allocation of scarce resources, forecasting the trajectory of the pandemic, and understanding the complex interactions between the pandemic and the intertwined social, political and economic systems of the 21st century. For example, pandemic spread can be stopped by having everyone stay home for an extended period of time (as was done in Wuhan), but this strategy is not feasible in open societies. Balancing the response with economic activity, social interactions, and political realities remains a significant challenge.

Foundations exist for the next generation of pandemic models. Compared to the coupled differential equations originally used to simulate how a virus spreads over time through homogenous populations, network models can easily incorporate more-realistic interactions among heterogeneous groups of people. Agent-based (networked) models work at an even finer grain, across time and space.

These sophisticated modeling approaches can be transformative: they can predict the course of the pandemic and the impact of various interventions, including testing and social distancing at different scales—although those predictions rapidly become outdated because of the effects of the interventions. Even more importantly, they are potentially more explainable, which is critical both to the public and to decision makers. Significant challenges remain, however:

- How to build models that are robust in the face of different assumptions, and that effectively communicate uncertainty in projections as assumptions are violated.
- How to gather and incorporate relevant data in real time to validate past predictions, update future projections and actively learn when modeling assumptions cause models to fail to capture real-world dynamics.
- How to model the sensing and monitoring systems that gather these data (e.g., testing and syndromic surveillance).
- How to understand and model the evolution of the pathogen in space and time, including its interaction with humans and their immune systems, as well as the effects of interventions and policies.

- How to rapidly incorporate the changing scientific understanding of the disease into the models and their underlying assumptions—and how to know when there is a need to do so.
- How to ensure that the models are sufficiently transparent and explainable so that the general population and policy makers understand how their actions and behaviors protect or expose themselves, their families, and their communities.
- How to incorporate socioeconomic factors such as occupation, race, income, and access to affordable health services that drive inequities in infection, hospitalization and death rates.
- How to effectively incorporate human behavior into the models, in the context of pandemic spread within economic and sociopolitical systems.
- How to validate these models. Traditional ways of using retrospective and predictive validity are rarely useful in crisis situations, when both the data about the situation, and our understanding of that situation, are limited.
- How to identify and balance possible tradeoffs between mitigation, privacy, security, and intellectual property in the face of public health emergencies.

Importantly, models should not focus narrowly on reducing epidemic spread, but rather be embedded in frameworks that target broader socio-economic impact. The models should also address growing trends in urbanization as well as climate change. They should not only incorporate known factors, but also reveal hidden variables and previously unknown factors—and support mitigations that reduce the risks of **all** causal variables.

This complex and important research agenda will require significant effort from the computing research community, in collaboration with experts from the social, behavioral, epidemic, and economic sciences, with sustained support at the national level and a strong, national-level infrastructure, as detailed in the Recommendations section at the end of this document. (See also the companion CCC Quad Paper entitled "The Rise of AI-Driven Simulators: Building a New Crystal Ball," which discusses the ways in which artificial intelligence strategies can aid in the modeling of complex problems).

**Infrastructure for Pandemic Informatics**
In a crisis, it is essential to have immediate access to enough computing power, data sources, and expertise to bring state-of-the-art methods to bear on the associated problems. In the current pandemic, the COVID-19 High Performance Computing (HPC) consortium has shown the value of advanced computing resources. However, this consortium was an ad hoc effort that required significant effort to establish and operate, and any delay can be critical in the context of a pandemic. Moreover, computing power is only one part of the solution. An even more fundamental need is information: comprehensive, clean data about all of the salient features of the situation, modified automatically and dynamically as that situation evolves, and distributed in a manner that preserves individual privacy. Computational infrastructure to manage data associated with vaccines, treatments, and long-term health effects of COVID-19 will be a particularly important element here. Finally, there is a need for experts in the applications, software, data, and system operations to be "on call" to minimize the barriers to entry for these resources.

Models and resources are only part of the landscape for pandemic preparation, resilience, and robustness, of course; research is also needed into privacy-preserving contact tracing; algorithms for optimal pooled testing; strategies to mitigate the effects of mis- and disinformation; understanding of the short- and long-term sociotechnical issues that arise when societies move to a distributed mode of life, learning, and work; and decision-support systems that help people make sense of the massive amounts of data from the world and from the models.

**Recommendations**

Bringing together the resources needed to support crisis response in the inevitable next pandemic, as well as, to carry out forward-looking research that will allow us to anticipate and perhaps divert its onset, would address many of these challenges thereby creating major benefits for the United States and the world. To that end, we recommend a multi-pronged approach that combines distributed and centralized resources, as well as, coordinated action from agencies, foundations, corporate research labs, academic departments, and national labs.

As a foundation for this, we recommend the creation of a distributed, pervasive *National Pandemic Informatics Infrastructure*. A dedicated *Pandemic Informatics Institute,* staffed on a full-time basis by a strong, interdisciplinary group of scientists, engineers, and support staff, would act as the nexus of this infrastructure, serving as a coordinator and central source of resources, best practices, etc. Critical resources—data sets, software, testbeds, and computational resources, as well as the essential expert support for accessing and using those resources—would be provisioned at various infrastructure sites across the nation to support both ongoing research regarding pandemics and the rapid ramp-up that accompanies these events. As highlighted above, pandemic response is a sociotechnical problem, and one in which policy plays a critical role. This has implications for the infrastructure: e.g., the development of privacy-preserving policies for rapid data sharing among various groups during crisis, or providing useful information to decision makers.

The research challenges outlined above fall at the boundaries between computing and other areas of work, notably the social sciences. The second element of our recommendations is a coordinated, multidisciplinary, multi-agency research effort to foster the necessary advances at these boundaries. Importantly, funding for this should not be diverted from existing agency budgets; rather, this demands wholly new funding streams. At the same time, agencies and foundations should continue the recently initiated programs that target pandemic informatics, extending them indefinitely and prescribing award levels and time horizons that allow for sustained work on these difficult problems. Importantly, all of these efforts must be deeply and fundamentally interdisciplinary, bringing together computing researchers with scientists, humanists, social scientists, and political scientists. A multi-agency initiative that brought together the NSF, DOE, DOD, CDC, IARPA, and the NIH, among others, would be useful in accomplishing this at a high level, but finer-grained policies and practices are also essential: e.g., requiring co-PIs from computing and the social sciences on a solicitation about models that incorporate human behavior.

*This material is based upon work supported by the National Science Foundation under Grant No. 1734706. Any opinions, findings, and conclusions or recommendations expressed in this material are those of the authors and do not necessarily reflect the views of the National Science Foundation.*

*For citation use: Bradley E., Marathe M., Moses M., Gropp W. & Lopresti D. (2020) Pandemic Informatics: Preparation, Robustness, and Resilience.*
*https://cra.org/ccc/resources/ccc-led-whitepapers/#2020-quadrennial-papers*

# Pandemic Informatics: Vaccine Distribution, Logistics, and Prioritization

**March 22, 2021 Addendum to Pandemic Informatics: Preparation, Robustness, and Resilience**

*Elizabeth Bradley (University of Colorado Boulder), Madhav Marathe (University of Virginia), Melanie Moses (The University of New Mexico), William D Gropp (University of Illinois Urbana-Champaign), and Daniel Lopresti (Lehigh University)*

When we authored our original quad paper in November 2020, we were writing from a distinct point at a unique moment in human history. Several months later, we find ourselves nearing what is hopefully the end of an exhausting marathon while facing a new and different set of challenges that are both surprising and yet also somehow predictable. There is much hope, but we must also acknowledge two key facts: (1) some parts of our society will be fortunate to make their way across the finish line while others still have many miles to go, and (2) the "once-in-a-lifetime" worldwide disruption caused by COVID-19 is likely to happen again, sooner than we wish to admit. With this in mind, and with the belief that computing research can play an important role now and in the future, we have decided to supplement our quad paper periodically with in-the-moment challenges where concerted effort now could produce solutions that would reduce suffering later. Each phase of the pandemic has had lessons to teach, and it seems prudent to do our best to document them as they arise so that we can learn from them.

In November, multiple companies announced successful trials of COVID vaccines. This was welcome news -- it told us that we finally had a pharmaceutical solution that can be used to manage the pandemic. While vaccine development was extremely successful, the distribution of vaccines has turned out to be tumultuous and chaotic. One major reason for this has been vaccine prioritization.

In an ideal world, one would have a single registry that would be used to track the end-to-end problem of how individuals are selected, how they are contacted, and how the vaccines are administered. This, however, is not how healthcare in the United States is administered, so it has to be done on a state-by-state basis. Each state manages its vaccine distribution through local constraints and objectives, which was led to a multitude of solutions.

The problem remains: how can we get vaccines in the arms of individuals efficiently? Efficiency is measured in multiple ways, (i) including providing clear, simple, and transparent methods for individuals to register, (ii) contacting individuals as vaccines start to arrive, (iii) ensuring that individuals get the vaccines based on the assigned priorities, (iv) ensuring that vaccines are not wasted, (v) ensuring that all eligible citizens are informed and have a chance to receive the vaccines, especially those from under-represented communities, and (vi) ensuring that misinformation about availability, hoarding, scalping, and side effects is reduced to the smallest extent possible.

The challenges in addressing the problems outlined above stem from several underlying structural issues:
- Lack of transparency in terms of vaccine production schedule and access to vaccination centers.
- Many citizens do not use the Internet or are not versatile in its use.
- Vaccine hesitancy stemming from lack of trust in the government, perception of side-effects, and related misinformation.
- One and two-dose vaccines, require to reserve vaccines in three/four weeks (based on the type) for individuals who got their first shot.
- The stringent cold chain requirement makes it challenging to store the vaccines for a long time and then use them within a specified period.
- Game-theoretic issues:
  - Providers and vaccine seekers try and find ways to get the vaccine.
  - Some people will never get the vaccine but should still benefit if others are vaccinated as herd immunity begins to increase.

These structural issues could be solved or addressed in the future through the following computer science research:
- Access to signups and status should be universally accessible and user-centered design should be employed. UX/UI and AI research could enable voice assistants such as Alexa and Siri to help achieve these goals.
- Vaccine signup websites do not scale to the traffic they are receiving. While this can be done because there are existence proofs (commercial online streaming services), it is hard to develop robust, scalable, universally accessible systems on a very short timescale. Research is needed to learn how to very quickly build, debug, and thoroughly test systems for the mass public.
- The rules for who is eligible vary widely and how they are coded in automated systems is unclear. This is an AI research problem: taking human-language specification of the priorities and the goals from public health experts as input, interpreting those in a way that assures high accuracy and fairness, and then continually monitoring the system to make sure it is achieving its goals and adjusting performance as necessary. Such systems should promote compliance with scientifically-based prioritization, but also enable flexibility, fairness, and speed in varied local contexts. These are FAIR (Findable, Accessible, Interoperable, Reusable) Principles, which put specific emphasis on enhancing the ability of machines to automatically find and use.
- Hundreds of thousands of municipalities are fielding their systems. Instead, it may make more sense to employ a few very high-quality community-developed and vetted open-source reference designs that are made freely available and customizable with minimal extra effort.
- Developing an information integration system that could be used to provide easy access to citizens as regards to availability, closest centers, priorities, etc. Some of this is now coming online by state health departments and CDC.

**See the initial November 2020 Paper [here](https://cra.org/ccc/wp-content/uploads/sites/2/2020/11/Pandemic-Informatics_-Preparation-Robustness-and-Resilience.pdf) ( https://cra.org/ccc/wp-content/uploads/sites/2/2020/11/Pandemic-Informatics_-Preparation-Robustness-and-Resilience.pdf ).**

*This addendum is part of a new monthly series of pandemic related addendums spurred by the continuing conversations of the Pandemic Informatics: Preparation, Robustness, and Resilience authors.*

*This material is based upon work supported by the National Science Foundation under Grant No. 1734706. Any opinions, findings, and conclusions or recommendations expressed in this material are those of the authors and do not necessarily reflect the views of the National Science Foundation.*

*For citation use: Bradley E., Marathe M., Moses M., Gropp W. & Lopresti D. (2021) Pandemic Informatics: Vaccine Distribution, Logistics, and Prioritization. https://cra.org/ccc/resources/ccc-led-whitepapers/#Pandemic-Informatics-Addendums*

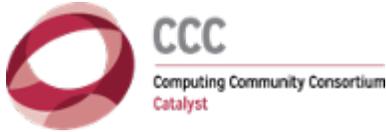

# Pandemic Informatics: Variants of Concern (VOC)
**April 22, 2021 Addendum to Pandemic Informatics: Preparation, Robustness, and Resilience**

*Elizabeth Bradley (University of Colorado Boulder), Madhav Marathe (University of Virginia), Melanie Moses (The University of New Mexico), William D Gropp (University of Illinois Urbana-Champaign), and Daniel Lopresti (Lehigh University)*

A year ago, few experts correctly predicted the toll the pandemic has now taken, nor the extraordinarily rapid development and administration of effective vaccines. A handful of wealthy countries have begun to curb COVID-19 infections and deaths through vaccination. The vaccine rollout has ramped up to cover half of all adults in the US with mRNA vaccines that are over 90% effective at preventing particularly severe disease, and also appear to greatly diminish transmission. Scientists have dramatically increased understanding of the SARS-CoV-2 virus, treatment, and vaccines. Yet, where the pandemic will be a year from now remains very difficult to predict, due in large part to rapidly spreading *variants of concern* (VOC).

Thus far, the three VOC spreading globally, and the two additional VOC recognized by the CDC in the US, have some combination of increased transmissibility, higher mortality, and/or ability to overcome immunity from prior infection or vaccination to varying degrees. These VOC present three challenges that computer science research can help address.

First, computing researchers, working in close collaborations with geneticists, immunologists, virologists, epidemiologists, and data scientists, can improve monitoring and tracking of VOC by developing methods to infer meaningful changes in epidemic dynamics from sparse and unevenly sampled data. The sequencing resources required to identify a small number of genetic changes in a 30,000 base pair genome are time-consuming, expensive, and often concentrated in high-resource places: not necessarily the locations where new variants will emerge. Additionally, sequencing is often concentrated on particular populations—i.e., health care workers and travelers—further complicating efforts to track how variants are spreading through the general population. Computer scientists can help develop optimization techniques for where, when, and how to selectively and effectively sample the effects of the evolving pandemic. This problem is complicated by the available technologies and associated costs but can be studied using ideas of robust optimization and location and estimation theory. Computational methods are needed to quickly identify globally when new variants are (a) spreading more rapidly than expected by chance or in relation to other variants, or (b) when the prevalence of a VOC is associated with increased mortality or hospitalization. These computational methods could help scientists to infer how much a particular increase in the viral spread is likely driven by a VOC versus social changes such as the loosing of restrictions or holiday gatherings.

Second, through collaboration with geneticists, immunologists, virologists, epidemiologists, and data scientists, computer scientists can help develop models of viral evolution and vaccine design to understand and, ideally, counter that evolution. Game-theoretic frameworks can aid in the understanding of the interplay between viral evolution and immunity by analyzing vaccination and prior infection. Anticipating which variants are likely to become dominant in the short and long term, based on estimates of current VOC prevalence and trends in viral evolution, can help in the design of boosters or wholly new vaccines that reduce infection, transmission, and severe disease. Modeling how the virus has, and may in the future, evolve—for example, in populations with uncontrolled transmission or in immunocompromised patients—may provide particularly useful insights into future VOC. Simple aggregated models or more detailed agent-based models can be developed by computing researchers to capture the interplay of immunity, viral evolution, and vaccine development.

Third, computing researchers, again working in close collaborations with scientists in other domains, can model treatments, the likely effect of vaccinations on transmission, and personal risk models for both vaccinated and unvaccinated people as new variants emerge. At a larger scale, computing research can contribute to risk models of different global vaccination programs. Currently, there is reason to hope that the US vaccination program will soon protect the majority of people who are most vulnerable to severe COVID-19 disease and death. However, that success could be short-lived if VOC evolved to evade the protection offered by those vaccinations. Modeling efforts are needed to understand how much risk is reduced, both globally and in the US, if worldwide vaccination were accelerated.

The past year should have taught us that apparent control of the virus can be temporary. Resurgences have been seen in wave after wave in place after place across the globe. We hope that the apparently successful efforts seen in the UK and Israel are harbingers of successful worldwide suppression of the virus, particularly in view of the nature of the more dangerous and more transmissible variant that is now dominant in the UK. However, when other variants (like those originating in Brazil, South Africa, and now possibly India) may evade prior immunity, the game may be changed. Models are essential to help us to understand where VOC emerge, how fast they spread, and how much they limit protection from vaccination or prior infection. Computer science can help with the rapid analysis given that uncertainty and the urgency of controlling VOC when they appear.

Much of the recent focus is on current VOC, which involves a handful of mutations that appear to increase viral transmission, infection, and/or immune escape. It is unknown whether this is a likely endpoint for the virus—i.e., whether continued evolution will lead to foreseeable changes—or whether this is simply the first generation of new variants that will open the door to further evolutionary possibilities. Computing research can help to address this uncertainty by developing models of how the virus may traverse the evolutionary landscape[1] and tools to quantify how evolutionary change may be driven by or affect the immune response [2].

---

[1] https://www.jstor.org/stable/j.ctt4cgcnc
[2] https://pubmed.ncbi.nlm.nih.gov/15218094/

These computational models and associated lab studies highlight that it is even harder to predict evolutionary trajectories than one might think.

Working closely with geneticists, immunologists, virologists, epidemiologists and data scientists computer scientists can (1) help to reduce the uncertainty about whether and how variants diminish the effectiveness of vaccination campaigns, (2) guide actions to limit VOCs that diminish effective vaccination, and (3) advance scientific understanding about the nature of the uncertainty of viral evolution.

**See the initial November 2020 Paper [here](https://cra.org/ccc/wp-content/uploads/sites/2/2020/11/Pandemic-Informatics_-Preparation-Robustness-and-Resilience.pdf) ( [https://cra.org/ccc/wp-content/uploads/sites/2/2020/11/Pandemic-Informatics_-Preparation-Robustness-and-Resilience.pdf](https://cra.org/ccc/wp-content/uploads/sites/2/2020/11/Pandemic-Informatics_-Preparation-Robustness-and-Resilience.pdf)).**

*This addendum is part of a new monthly series of pandemic related addendums spurred by the continuing conversations of the Pandemic Informatics: Preparation, Robustness, and Resilience authors.*

*This material is based upon work supported by the National Science Foundation under Grant No. 1734706. Any opinions, findings, and conclusions or recommendations expressed in this material are those of the authors and do not necessarily reflect the views of the National Science Foundation.*

*For citation use: Bradley E., Marathe M., Moses M., Gropp W., Lopresti D. (2021) Pandemic Informatics: Pandemic Informatics: Variants of Concern (VOC). [https://cra.org/ccc/resources/ccc-led-whitepapers/#Pandemic-Informatics-Addendums](https://cra.org/ccc/resources/ccc-led-whitepapers/#Pandemic-Informatics-Addendums)*